\documentclass[sigconf]{acmart}
\usepackage{caption}
\usepackage{subcaption}
\usepackage{hyperref}

\copyrightyear{2024}
\acmYear{2024}
\setcopyright{rightsretained}
\acmConference[SIGIR '24]{Proceedings of the 47th International ACM SIGIR Conference on Research and Development in Information Retrieval}{July 14--18, 2024}{Washington, DC, USA}
\acmBooktitle{Proceedings of the 47th International ACM SIGIR Conference on Research and Development in Information Retrieval (SIGIR '24), July 14--18, 2024, Washington, DC, USA}
\acmDOI{10.1145/3626772.3657655}
\acmISBN{979-8-4007-0431-4/24/07}
\begin{document}

\title{GOLF: Goal-Oriented Long-term liFe tasks supported by human-AI collaboration}

\author{Ben Wang}
\email{benw@ou.edu}
\orcid{0001-8612-1185}
\affiliation{%
  \institution{University of Oklahoma}
  \city{Norman}
  \state{Oklahoma}
  \country{USA}
}

\renewcommand{\shortauthors}{Wang.}

\begin{abstract}
The advent of ChatGPT and similar large language models (LLMs) has revolutionized the human-AI interaction and information-seeking process. Leveraging LLMs as an alternative to search engines, users can now access summarized information tailored to their queries, significantly reducing the cognitive load associated with navigating vast information resources. This shift underscores the potential of LLMs in redefining information access paradigms. Drawing on the foundation of task-focused information retrieval and LLMs' task planning ability, this research extends the scope of LLM capabilities beyond routine task automation to support users in navigating long-term and significant life tasks. It introduces the GOLF framework (Goal-Oriented Long-term liFe tasks), which focuses on enhancing LLMs' ability to assist in significant life decisions through goal orientation and long-term planning. The methodology encompasses a comprehensive simulation study to test the framework's efficacy, followed by model and human evaluations to develop a dataset benchmark for long-term life tasks, and experiments across different models and settings. By shifting the focus from short-term tasks to the broader spectrum of long-term life goals, this research underscores the transformative potential of LLMs in enhancing human decision-making processes and task management, marking a significant step forward in the evolution of human-AI collaboration.
\end{abstract}




\maketitle
© Wang et al., 2024. This is the author's version of the extended abstract of his dissertation work. It is posted here for your personal use. Not for redistribution. The one-page definitive Version of Record was published in ACM SIGIR 2024, https://doi.org/10.1145/3626772.3657655.

\section{Motivation}
The release of ChatGPT has sparked considerable interest in the interaction between humans and Artificial Intelligence (AI). People use ChatGPT and other large language model (LLM) tools for various purposes. One prevalent usage is as an alternative to a search engine \cite{bahrini2023chatgpt, skjuve2023user}. These LLM tools can summarize information and adjust the response according to people’s questions, mitigating the cognitive load of human users when browsing and processing massive information using a traditional search engine \cite{skjuve2023user, jin2023retrieve}. This shift underscores the potential of LLMs and calls for a new paradigm of information access \cite{shah2024envisioning}.

Building upon the foundational theories of interactive information retrieval (IIR), such as berry-picking \cite{bates1989design} and information foraging \cite{pirolli1999information}, this work introduces a novel paradigm for task-oriented information seeking that leverages the collaborative potential of human-AI interactions. This study articulates the paradigm through an inventive analogy, likening the process of LLM-based information access to playing golf. In this analogy, the user, akin to a golfer, fine-tunes their strategies (or prompts) to efficiently achieve their objective (satisfying their information need or completing a task). This model shifts the emphasis from the traditional, often complex search processes examined in IIR to a more streamlined, outcome-focused approach.

By focusing on the strategic use of prompts and prioritizing the end goal, the LLM-based information access system addresses the common challenges of navigating overwhelming information and reduces the cognitive burden associated with conventional search. It offers a more intuitive and direct pathway to task completion, making the process of information retrieval not only more efficient but also more adaptively aligned with the user's needs. Through this lens, my study highlights how LLMs can transform information access, moving beyond traditional IIR to a more goal-centric, user-friendly interaction system.

LLMs become capable of executing routine tasks with minimal human input and have developed abilities to break down complex objectives into manageable subtasks and utilize external tools for comprehensive problem-solving and task automation with minimal human input \cite{brown2020language, ouyang2022training, huang2024understanding, yao2022react}. However, the question arises: what further benefits can humans derive from deeper collaboration with LLMs, especially in navigating long-term and significant life tasks? The significant challenges in IIR still persist for LLMs, particularly in interpreting dynamic user intents, managing ambiguous instructions, and ensuring alignment with human values over extended interactions \cite{azzopardi2021cognitive, liu2020identifying, liu2022leveraging, wang2024task}. Therefore, this research aims to bridge the gap by exploring how LLMs can support users in accomplishing long-term task goals and making life-significant decisions. Specifically, this study proposes the GOLF framework: Goal-Oriented Long-term liFe tasks supported by human-AI collaboration. By emphasizing goal orientation and long-term planning, the GOLF framework aims to explore and enhance LLMs' abilities to support significant life decisions and address challenges of long-term task management.

\section{Related Work}
\subsection{Tasks in Information Seeking and Retrieval}
In the context of Information Seeking and Retrieval (ISR), tasks are multifaceted activities that can range from everyday queries to complex professional information needs \cite{vakkari2003task, xie2009dimensions}. These tasks often trigger a cascade of information-focused actions such as searching, acquiring, organizing, synthesizing, and disseminating information to achieve specific goals \cite{bystrom2005conceptual, jarvelin2015task}. The process of addressing these tasks frequently involves navigating through the Information Search Process (ISP), a conceptual model that outlines various stages ranging from initiation, exploration, selection, formulation, and collection, to the presentation of information \cite{kuhlthau1991inside}. This model underscores the dynamic nature of information seeking, spotlighting how individuals' approaches evolve as they gain clarity and refine their information needs. Task hierarchies in ISR further delineate how tasks are structured and interrelated, positing that complex work tasks may encompass nested sub-tasks related to information seeking and searching \cite{toms2019information, bystrom2005conceptual}. Building on these insights, \citet{soufan2021untangling} integrated task taxonomy proposed in previous studies and sought to harmonize disparate understandings of tasks within ISR. They incorporated considerations of task levels from work roles, activities, tasks, to subtasks, thereby offering a comprehensive framework that captures the hierarchical nature within various work and real-life contexts \cite{bystrom2005conceptual, kuhlthau1991inside, toms2019information}.

Recent progress in task support research has expanded the scope and efficiency of IR systems. A significant focus has been on understanding and enhancing user-task interactions within search environments. \citet{shah2023taking} emphasized the need to focus on user tasks rather than just query processing, proposing a new framework for task-based Information Retrieval to guide future research. \citet{hassan2014supporting} introduced methods to automatically identify and recommend subtasks using a task graph derived from search logs, facilitating users in navigating complex search tasks. \citet{hassan2012task} extended support by generating task tours that map out necessary steps for complex tasks, improving users' search efficiency. \cite{song2016query} presented a proactive approach by predicting and recommending repeated tasks using a novel deep learning framework, pushing towards a future of query-less search engines. \citet{white2019task} focused on optimizing task management systems through machine learning models that automatically detect task completion, enhancing digital assistant functionalities. Furthermore, \cite{liu2020identifying} investigates the dynamic nature of complex search tasks and reveals the nonlinear transitions of task states. These studies emphasize the importance of task understanding, prediction, and support, thus laying a robust foundation for future innovations in task-based information systems.

\subsection{LLM Agent and Task Automation}
LLMs represent advanced capabilities centered around natural language understanding and generation, which enables them to execute a diverse array of human tasks. The capacity of LLMs is heavily reliant on their pre-training on extensive datasets, enabling a wide range of generalization across tasks \cite{brown2020language, ouyang2022training}. The concept of an LLM agent denotes an entity capable of performing intentional actions and encompassing autonomy, reactivity, proactiveness, and social ability \cite{schlosser2015agency, park2023generative}. LLM agents are designed to facilitate everyday tasks and improve task efficiency while reducing user workloads \cite{li2023camel, xi2023rise}. The inherent ability of LLM agents to understand, decompose, plan, and subsequently complete tasks is fundamentally attributed to their emergent reasoning and planning capacities, exemplified through strategies like Chain-of-Thought (CoT) and problem decomposition \cite{wei2022chain, yao2022react}. 

Task planning and automation emerge as key functionalities of LLM agents. They can decompose complex tasks into manageable sub-tasks, plan sequences of actions, and interactively explore environments to achieve objectives, often leveraging reinforcement learning to mimic human behavior \cite{huang2024understanding, park2023generative}. This is critically enhanced by their natural language understanding, enabling them to process instructions and execute tasks that require common-sense knowledge and adapt to dynamic situations \cite{li2023camel}. The sophisticated integration of LLM capabilities with task-specific strategies facilitates effective task completion and automation, thereby offering practical value in enhancing efficiency and accessibility for users \cite{wu2023autogen}.

\section{Research Questions}
This research proposes a forward-looking approach, shifting focus from the short-term tasks that AI agents can autonomously complete to the long-term tasks related to human life decisions and development. Here, short-term tasks refer to smaller-scale and routine tasks that LLM agents can automate with less human intervention. The long-term tasks encompass broader personal life goals or development in aspects like health, finances, education, and professional development. These tasks cannot be fully completed by LLM agents alone but require significant human involvement. Nonetheless, LLM agents can still assist human users to break down a complex task into many steps and reduce human workloads in short-term subtasks. Therefore, this research aims to explore how LLM agents can aid users in long-term life tasks and decisions, leveraging task planning and automation capabilities. It emphasizes the synergistic potential of human-AI collaboration, where the goal is not to create AI systems that replace humans but to enhance human abilities, enrich lives, and address complex challenges more effectively. To address this, it is important to answer these research questions (RQs): 
\begin{itemize}
    \item[RQ]1. What is the potential framework of AI agents that supports long-term human tasks? 
    \item[RQ]2. What are users’ needs in completing long-term tasks with collaborating with AI agents? 
    \item[RQ]3. How to deploy and evaluate LLM systems that support long-term human tasks?
\end{itemize}

\section{Proposed Methodology}
This section outlines the methodology, focusing on the implementation, evaluation, and application of the GOLF framework. It begins by introducing the GOLF framework, and detailing its task structure and process. Following this, I describe the implementation process, which involves a comprehensive simulation study designed to test the framework's efficacy under varied conditions. The evaluation involves model and human assessments for developing a dataset benchmark specifically tailored to long-term life tasks. Finally, my study explores the application of the GOLF framework across different models, conducting experiments to ascertain its versatility and effectiveness in diverse settings. Through this multifaceted methodology, my study aims to provide a thorough examination of the GOLF framework's potential in how LLMs can assist human users with long-term life tasks.

\subsection{The GOLF Framework}
\begin{figure*}
\centering
\begin{subfigure}{0.33\linewidth}
  \centering
  \includegraphics[width=\linewidth]{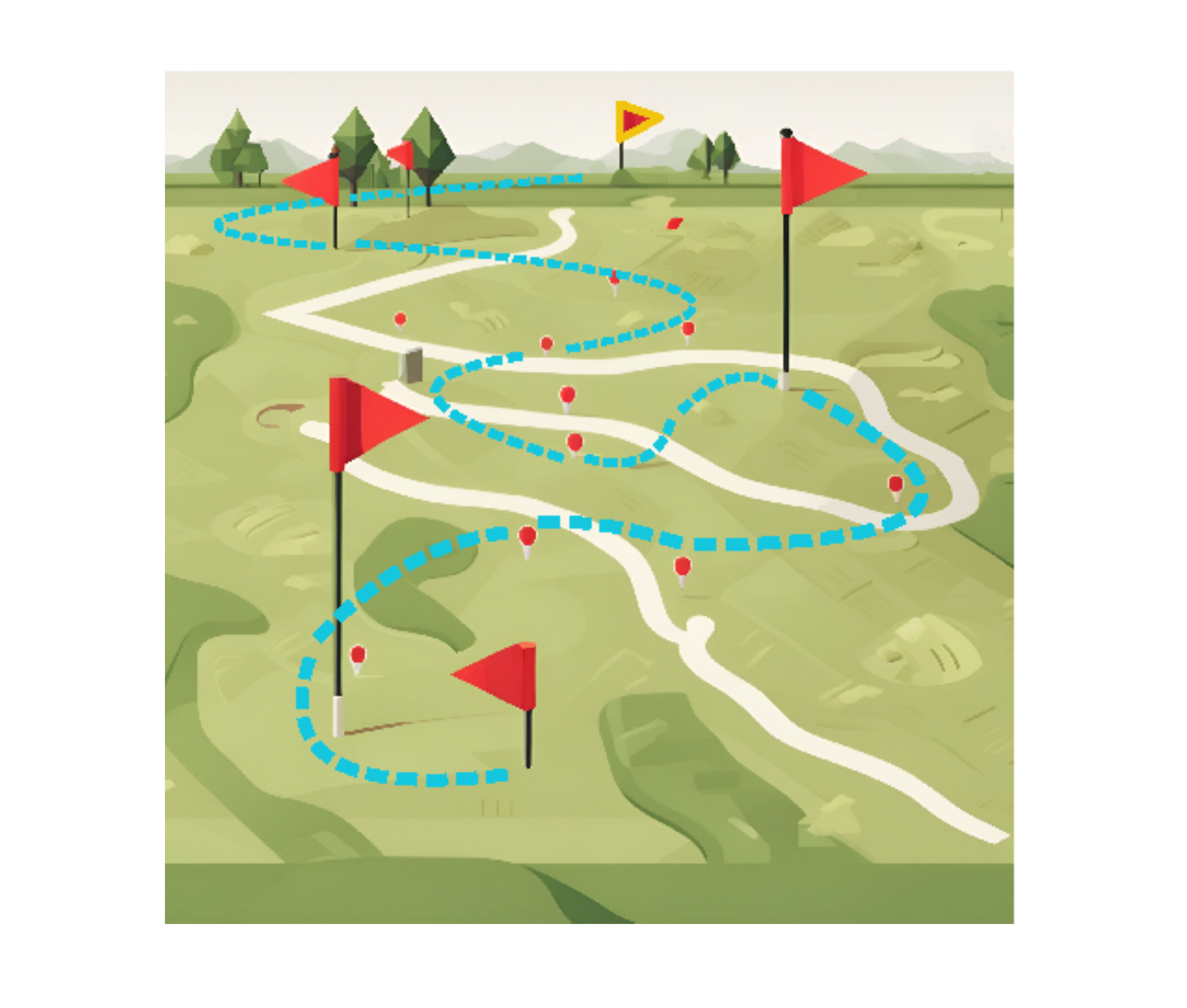} 
  \vspace*{-6mm}
  \caption{}
  \label{course}
\end{subfigure}
\begin{subfigure}{0.33\linewidth}
  \centering
  \includegraphics[width=\linewidth]{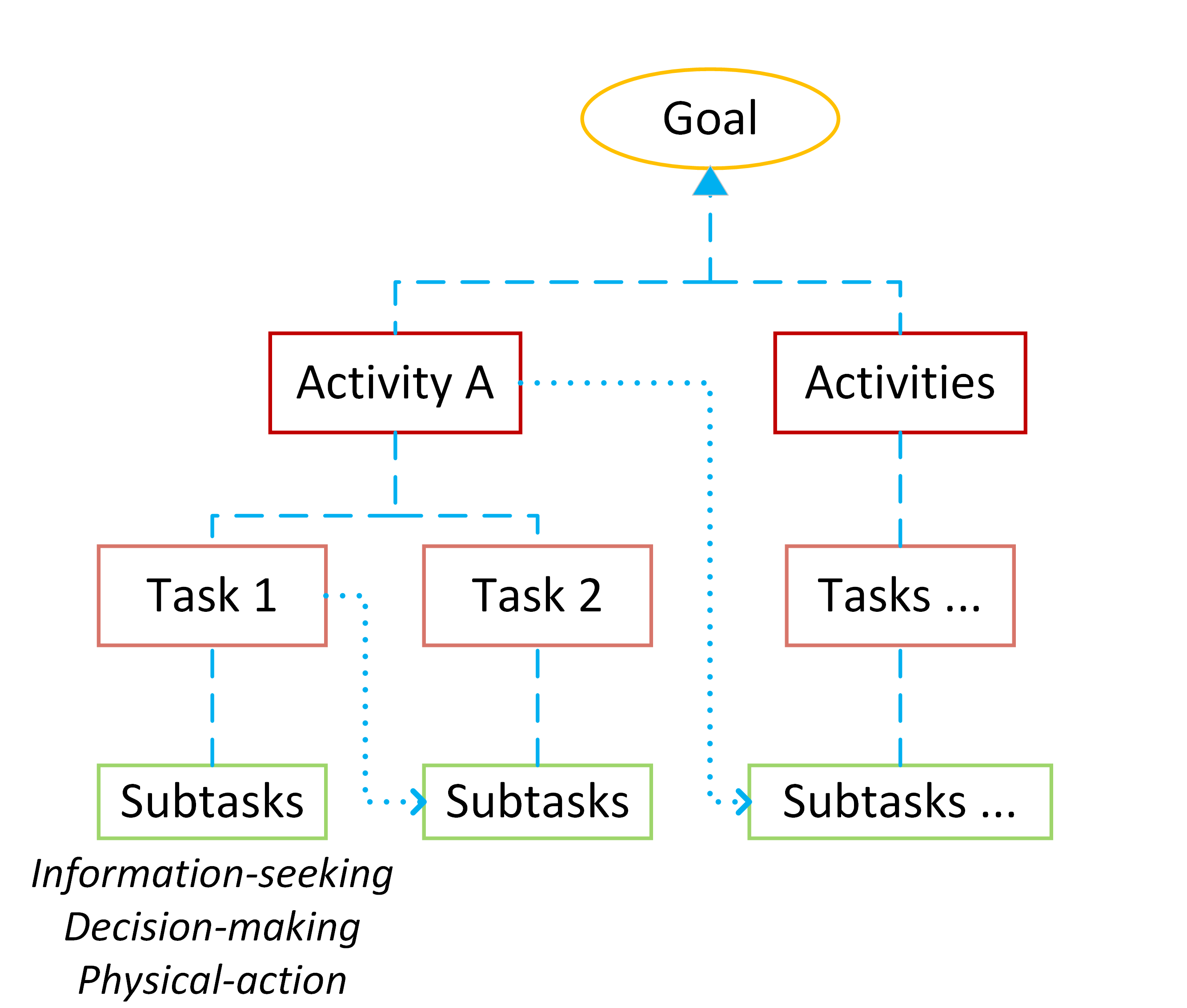}
  \vspace*{-6mm}
  \caption{}
  \label{task}
\end{subfigure}
\begin{subfigure}{0.33\linewidth}
  \centering
  \includegraphics[width=\linewidth]{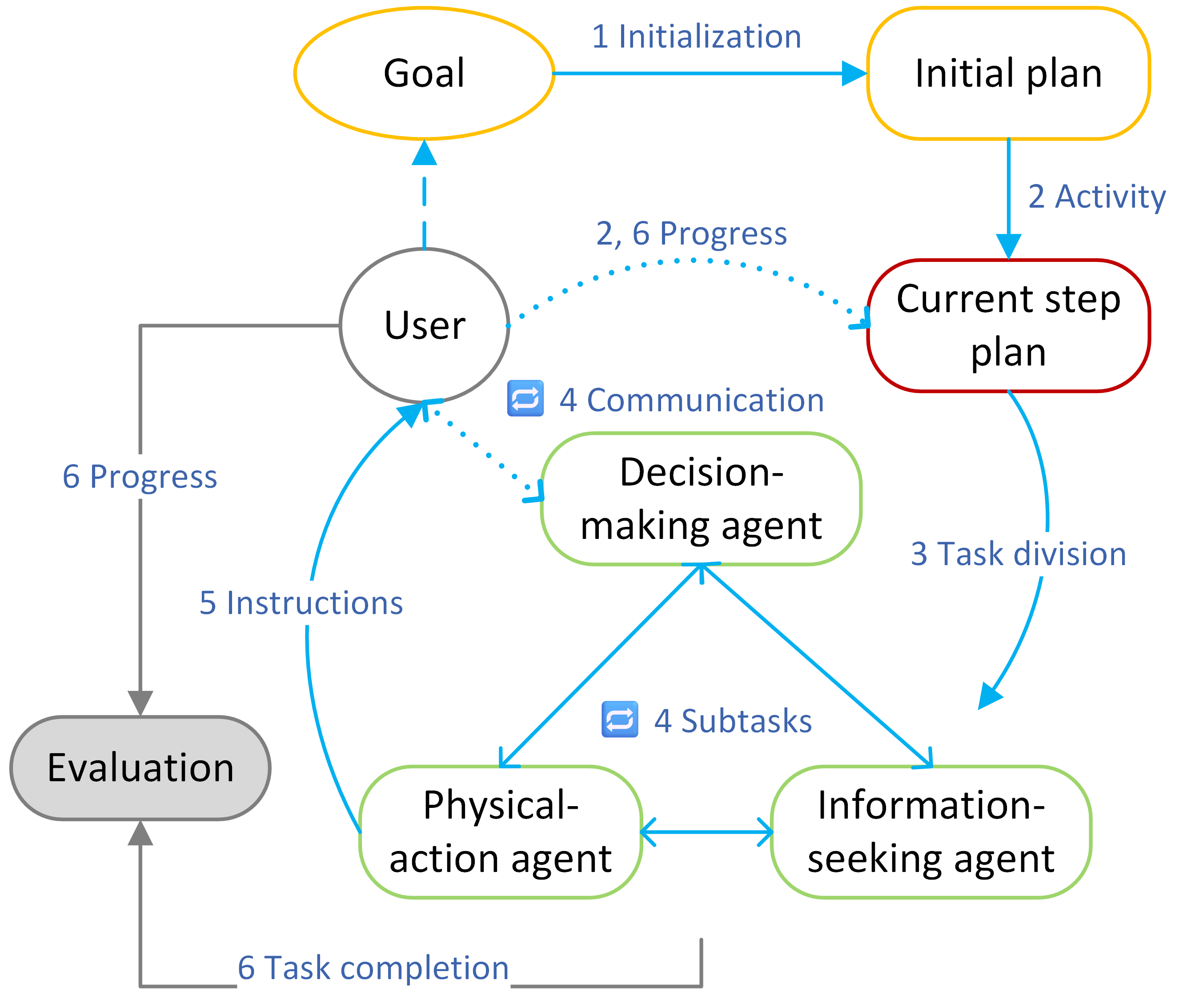}
  \vspace*{-6mm}
  \caption{}
  \label{framework}
\end{subfigure}
\vspace*{-3mm}
\caption{The GOLF framework: (a) a simplified illustration of a golf course (the background is generated by StableDiffusion); (b) a variation of the task taxonomy \cite{soufan2021untangling}; (c) the task process in the GOLF framework. The blue lines represent the potential path to complete the task, activity, or goal. Dotted lines represent users' actions, and solid lines represent LLM agents' actions.}
\label{golf}
\end{figure*}

Figure \ref{golf} presents the GOLF framework, which includes a task taxonomy and outlines the process for task management under this framework. To elucidate the GOLF framework's structure and approach, Figure \ref{course} depicts an analogy with the game of golf. In golf, players aim to sink the ball into a series of holes on the course with the fewest strokes possible, navigating recommended paths (blue dashed lines) and overcoming the challenge of landing spots (red pointers) between flags. Similarly, the GOLF framework envisions the completion of complex tasks as a strategic journey towards a final goal, incorporating a sequence of activities, tasks, and subtasks. This task taxonomy in Figure \ref{task} decomposes complex human tasks from the information-seeking aspects \cite{soufan2021untangling}, while the GOLF framework considers three types of subtasks: information-seeking, decision-making, and physical-action tasks, based on human information processing theory and task management principles.

Figure \ref{framework} illustrates the task process within the GOLF framework, which operates on a sophisticated multi-agent system designed to facilitate user support and workload distribution for achieving long-term goals. This system involves multiple LLM agents, including an initial planner, a step planner, task-specific agents, and an evaluator. By implementing prompt engineering with human task management principles, including "Getting Things Done: Capture, Clarify, Organize, Reflect, Engage" \cite{allen2015getting} and "PDCA: Plan, Do, Check, Adjust" \cite{moen2006evolution}, LLM agents process the task as follows:
\begin{itemize}
\item[1.] Initial Planning: The user inputs a brief goal statement, prompting the initial planner to draft an abstract plan outlining broad goals, activities, tasks, and subtasks.
\item[2.] Step Planning: Based on the initial plan and the user's progress, the step planner extracts a step plan that specifies the current activity for completion.

\item[3.] Task Assignment: This step plan is then segmented into information-seeking, decision-making, and physical-action tasks, and assigned to corresponding LLM agents for execution.
\end{itemize}

Specifically, the information-seeking task requires the agent to help the user seek, understand, and process objective information from the web. The decision-making task requires the agent to communicate with the user to collect additional personal information, preferences, and considerations, to help the user make judgments, evaluations, and decisions. The physical-action task requires the agent to instruct the user to make actionable steps or interactions with the real world to proceed with the task. This task distribution aims to reflect assistance through efficient information processing, informative decision-making, and realistic planning.

\begin{itemize}
\item[4.] Multi-Agent Coordination: Agents collaboratively generate actionable plans by integrating external information with user-provided data, ensuring the plan remains aligned with the user's needs while minimizing information overload.

\item[5.] User Engagement: The user follows the actionable plan, providing progress updates and feedback.

\item[6.] Evaluation and Iteration: The step planner refines subsequent step plans according to user progress. The task completion of the three agents and user progress are combined to assess the effectiveness in relation to both the overarching goal and the specific subtask plans.
\end{itemize}

Steps 2-6 are iterative as the process continues until the user completes the final goal. The step planner generates revised plans for subsequent steps based on the updated user states and task progress. These updated plans undergo another cycle of collaboration by the three agents.

Throughout this process, the GOLF framework emphasizes seamless integration between user inputs and agent assistance, ensuring that plans are dynamically updated to reflect the user's changing needs and circumstances. This iterative, multi-agent approach aims to distribute the workload effectively and adapts to the user's evolving preferences and the task's requirements, considering a personalized and efficient path to achieving long-term objectives.

\subsection{Simulation Study}
To test the interaction processes within the multi-agent framework, this study employs Autogen\footnote{https://microsoft.github.io/autogen/} \cite{wu2023autogen}, a tool designed for developing and managing multiple LLM agents that simulate roles within user-system interactions. The implementation needs to facilitate the creation of realistic simulation environments, enabling the study to explore a variety of long-term life tasks across different domains such as health, finance, education, and professional development. For example, in the health domain, this study explores simulated scenarios like chronic disease management, where users engage with an AI-provided management plan to navigate their recovery process. In the financial domain, tasks such as buying a house are modeled to require careful consideration of constraints and sequential steps for successful completion. Each simulation incorporates task scenarios with consistent final goals but allows for variable interactions and outcomes based on simulated user characteristics, such as knowledge level, AI acceptance, and engagement with specific subtasks. This approach enables the study to introduce controlled variations in user behavior and preferences, directly influencing task progress and the effectiveness of goal achievement.

The simulation study will run on GPT-4-turbo and start from a basic and ideal scenario and then introduce complexity into the simulation. In the basic scenario, agent collaboration follows a predetermined task sequence without real-time updates, and more sophisticated conditions incorporate dynamic updates based on user feedback and changing environmental factors. The collaboration among the three agents is dynamically adjusted to reflect ongoing communication with the user and new information gathered by the information-seeking agent. These conditions allow the study to examine the framework's adaptability to varying levels of user engagement, agent collaboration, and external information inputs.

\subsection{Evaluation}
To assess the efficacy of the GOLF simulation framework and the completion of tasks by simulated users under various conditions, the evaluation employs a two-fold approach involving an LLM evaluator and human assessors. Initially, the LLM evaluator will analyze the effectiveness of the three-agent collaboration in achieving task completion, taking into account user characteristics and task progress. This evaluation will measure both the achievement of the final goal and the efficiency of the step-wise task process, utilizing predefined criteria to ensure a comprehensive assessment. Subsequently, the simulation outcomes will be evaluated by human assessors via crowdsourcing. This phase involves presenting detailed scenarios and user profiles to the assessors, who will then evaluate the effectiveness of the agent collaboration, including components such as user communication, integration of external information, and the viability of the action plans generated.

In addition to the evaluation of the simulated data, this study will also examine the practical value of the GOKF framework with user human users. The following user study will extend invitations to individuals from the initial crowdsourcing pool as well as from other sources, such as college mailing lists and LinkedIn, to ensure a diverse range of user groups and long-term task topics. These participants will be encouraged to specify their personal long-term objectives and current progress, providing a real-world testbed for evaluating the GOLF framework's plan generation capabilities. Feedback from these engagements, encompassing both simulated and actual task scenarios, will inform the development of evaluation metrics. These metrics will underpin the creation of a comprehensive benchmark designed to enhance the training and assessment of LLM agents in facilitating long-term task completion.

\subsection{Model Experiment and Deployment}
The initial simulation exclusively leverages the GPT-4 model, known for its effectiveness and state-of-the-art performance, but it still takes relatively high computational costs. The next phase aims to incorporate and train open-source models. This expansion seeks not only to evaluate the GOLF framework's adaptability across different AI architectures but also to ensure its accessibility and efficiency in real-world applications, particularly for long-term task assistance. Considering the deployment context is crucial, especially when the end goal is to embed the GOLF framework within daily life applications accessible on various platforms, including mobile devices. To this end, models with fewer parameters, such as Llama2 7B and Phi-2, emerge as viable alternatives. These models may offer a balance between computational efficiency and the cognitive depth required for complex task management, making them suitable for on-device deployment where resources are limited.

The training and evaluation phases will utilize the dataset curated from the simulation study and the benchmark, specifically designed to capture the nuanced requirements of long-term task planning and execution within the GOLF framework. Deployment considerations extend beyond model selection to include the functional capabilities of the agents, particularly the information-seeking agent's ability to access and utilize external tools and resources. This capability is paramount for maintaining the framework's efficacy in providing up-to-date, contextually relevant information to users.

Finally, the deployment phase will incorporate an online testing component, enabling real-time collection of user feedback. This critical step not only serves to validate the framework's performance in live scenarios but also establishes a human-in-the-loop process for continuous improvement. Through this iterative process, user insights and experiences will directly inform further refinements, ensuring the GOLF framework is aligned with user needs and preferences for long-term goals.

\section{Specific Research Issues for Discussion}
In the concluding section of my proposal, I explore critical challenges and limitations that may trigger meaningful discussions. illuminate the intricate relationship between human users and LLM agents in managing long-term tasks collaboratively. In the proposed framework and simulation study, diverse factors are set to influence the interaction. The factors include task management strategies, simulated user differences, and simulated environment changes. However, the simulation effectiveness depends on the LLMs' reasoning performance, which might be insufficient to reflect real-life conditions. Subsequently, a significant methodological challenge arises in designing user studies for long-term tasks within the constraints of a doctoral research timeline. Therefore, this study proposes a hybrid methodology that merges simulation studies with targeted user studies on certain activities or subtasks as segments of long-term tasks. This approach seeks to obtain intermittent feedback on the framework's task plans and to uncover partial effectiveness for enhancing long-term task collaboration.

This research further prompts questions related to the shift in interactive and conversational IR brought by LLM-based conversational AI. Understanding how these differences affect user interactions will inform the implementation of the proposed framework and the simulation. Moreover, evaluating the proposed framework or general LLM tools for long-term tasks requires a thorough assessment beyond task completion efficiency and user progress. It must encompass AI acceptance, ethical considerations, and privacy implications. Establishing an evaluation framework that addresses these dimensions is crucial for ensuring that LLMs' support for long-term tasks adheres to ethical standards and safeguards user privacy. Such a framework and general LLM tools are supposed to nurture a trust-based and ethically responsible human-AI collaboration society.

\begin{acks}
This work is supported by the National Science Foundation (NSF) Award IIS-2106152, a grant from the Seed Funding Program of the Data Institute for Societal Challenges, the University of Oklahoma, and a fund from Microsoft for Startups Founders Hub. Any opinions, findings, conclusions or recommendations expressed in this material are those of the authors and do not necessarily reflect those of the sponsors. The golf analogy was inspired by Dr. Andrew Fagg. This work is conducted under the supervision of Dr. Jiqun Liu.
\end{acks}

\bibliographystyle{ACM-Reference-Format}
\bibliography{sample-base}


\end{document}